\documentclass[11pt,a4paper]{article}
\usepackage{jheppub}

\usepackage{amssymb,amsfonts,amsmath,graphicx,}
\usepackage{color}
\usepackage{enumerate}
\usepackage{braket}
\usepackage{epstopdf}

\usepackage{varioref}
\usepackage{slashed}

\title{Neutrino Transition in Dark Matter}

\author{Eung Jin Chun}

\affiliation{Korea Institute for Advanced Study, Seoul 130-722, Korea}

\emailAdd{ejchun@kias.re.kr}

\preprint{KIAS-P21056}

\abstract{
An ultralight dark matter may have interesting implications in neutrino physics which have been studied actively in recent years.  It is pointed out that  there appears yet unexplored medium effect in neutrino transitions which occurs at the first order in perturbation of the neutrino-medium interaction. We derive the general formula for the neutrino transition probability in a medium which describes the standard neutrino oscillation as well as the new medium contribution.  It turns out that such an effect constrains the model parameter space more than ever.
}

\begin{document}
\maketitle

\section{Introduction}

An ultralight boson is an attractive dark matter candidate which can behave like a classical wave \cite{Hu:2000ke,Hui:2016ltb}.
Its wavelength could be on astrophysical scales and thus leave observable effects in galactic dynamics \cite{Hayashi:2021xxu}.
As the classical field value of such a dark matter may be as large as $\phi_0 \sim 10^{11}\mbox{GeV}(10^{-22}\mbox{eV}/m_\phi)$ GeV,
it can also alter the Standard Model dynamics when it couples to quarks and leptons. In recent years, its relevance to neutrino physics has been explored extensively [4-21] as the neutrino sector would require some new physics for the origin of tiny neutrino masses. 
The ultralight dark matter coupling to neutrinos  modifies the standard neutrino oscillations in various ways and thus may hint at some anomalous phenomena in observations. More dramatically, the observed neutrino oscillations would be solely due to the medium effect  changing the dispersion relation of neutrinos propagating in a medium \cite{Weldon:1982bn,Nieves:1989ez}, which was suggested originally by Wolfenstein  \cite{Wolfenstein:1977ue}.

In this article, we discuss  yet unexplored phenomenon of neutrino transitions through absorbing/emitting tiny momenta from/to the medium. Unlike the standard neutrino oscillations caused by mass-squared differences in vacuum or in medium through modified dispersion relations \cite{Choi:2019zxy,Choi:2020ydp}, it occurs at the first order in perturbation of the medium-neutrino interaction. 
This neutrino transition could modify  the standard neutrino oscillations which have been established firmly, and thus is limited by various neutrino observations. Considering the  standard picture that neutrino oscillations are due to tree-level neutrino masses and mixing, we will obtain constraints on the neutrino coupling to the scalar dark matter $\phi$ in terms of its mass $m_\phi$.  
We will also consider the opposite case that the observed neutrino oscillations are mainly due to the medium effect, and describe the limits on the bare mass of a neutrino or a mediator fermion.

\medskip

\begin{figure}
\centering
\includegraphics[width=0.9\textwidth]{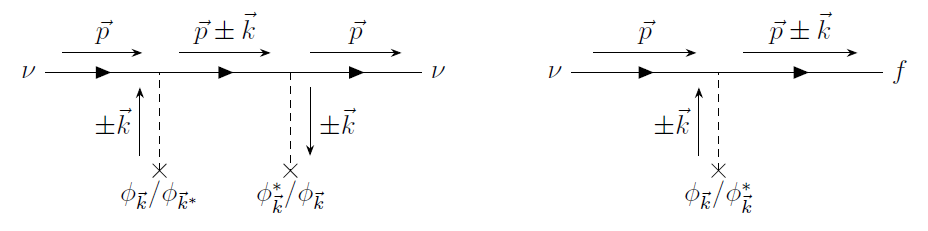}
\caption{(Left) The neutrino forward scattering in a coherent scalar background that modifies  the dispersion relation with an additional contribution of $|g\phi_{\vec{k}}|^2$. (Right) The neutrino transition processes $\nu(\vec{p}) \to f(\vec{p}\pm \vec{k})$ occuring in the background of $\phi_{\vec{k}}/\phi^*_{\vec{k}}$. }
\label{fig:nutr}
\end{figure}

The best-known way of generating cold but light dark matter is the misalignment mechanism \cite{Preskill:1982cy, Abbott:1982af,Dine:1982ah} in which an initial field value displaced from the true minimum drives the coherent oscillation behaving like pressureless dark matter.  When considering such a dark matter background,  it is useful  to remind us of some features of a classical field as a coherent quantum state \cite{sudarshan,glauber}.  Some introductory discussions can be found in Refs.~\cite{bergstrom, itzykson}. 
For a real scalar quantum field, $\hat \phi(x) =\int\!{d^3\vec{p} \over (2\pi)^3 2E_{\vec{p}} }[a_{\vec{k}} e^{-ik\cdot x} + a^\dagger_{\vec{k}} e^{ik\cdot x}]$, its coherent state is described by $|\phi_c\rangle \propto e^{\int_{\vec{k}} [\phi_{\vec{k}} a_{\vec{k}}^\dagger -
{\phi}^*_{\vec{k}} a_{\vec{k}}]}|0\rangle$ which is an eigenstate of the annihilation and creation operators: $\langle \phi_c| \hat a_{\vec{k}} |\phi_c\rangle = \phi_{\vec{k}}$ and  $\langle \phi_c| \hat a^\dagger_{\vec{k}} |\phi_c\rangle = \phi^*_{\vec{k}}$.
Such a non-vanishing field value in the background can lead to important effects on neutrinos propagating through the medium.
As is well-known, the coherent forward scattering that leaves the medium unperturbed (depicted in the left panel of Fig.~\ref{fig:nutr}) amounts to an additional contribution to the neutrino propagator  and thus modifies the dispersion relation. 
As a result, the neutrinos can receive the medium-induced mass-squared proportional to $|g\phi_{\vec{k}}|^2$ \cite{Choi:2019zxy,Choi:2020ydp}.
Remarkably there could appear a novel process of transition from an initial state $\nu(\vec{p})$ to two different final states $f(\vec{p}\pm \vec{k})$ as shown in the right panel of  Fig.~\ref{fig:nutr}. The transition amplitudes are proportional to the background field value $g \phi_{\vec{k}}$ and $ g\phi^*_{\vec{k}}$,
 and thus resulting transition probabilities are also proportional to $|g\phi_{\vec{k}}|^2$.

\medskip

For the clarity of our discussion, let us consider a monochromatic state 
$\phi_{\vec{k}} =(2\pi)^3 2 E_{\vec{k}}\, $ $\delta^3(\vec{k}-\vec{k}_0)\, \phi_0 $
for which  the classical field $\phi_c(x)$, which is an expectation value of the quanum field $\hat\phi(x)$ in the medium,
 takes the simple form:
\begin{equation} \label{phic}
 \phi_c(x) =
\langle \phi_c| \hat\phi |\phi_c\rangle= \phi_0 e^{-i k_0\cdot x} + \phi_0^* e^{+i k_0\cdot x}
\end{equation}
where $k_0=(E_{\vec{k_0}}, \vec{k}_0)$ and $\phi_0$ is a  complex number.   
 Our calculation can be generalized to the complext scalar quantum field where $a^\dagger$ is replaced by an independent creation operator $b^\dagger$, and  $\phi_0^*$ replaced by an independent complex number $\bar\phi_0$. 
In the background of dark matter, we have approximately $k_0 \approx (m_\phi, m_\phi \vec{v}_{\phi})$ with $v_{\phi}\approx 10^{-3}$ and thus we can safely put $\vec{k}_0= 0$ in the final results. 
 Furthermore, the relation 2$|\phi_0|^2 = \rho_{\phi}/m^2_\phi$ holds taking $\rho_{\phi}=0.3 \mbox{GeV}/\mbox{cm}^3$ for the local dark matter density. 
For the realistic situation, one needs to consider a velocity dispribution $|\phi_{\vec{k}}|^2$ of dark matter centered around 
$\vec{k}=\vec{k}_0$.  It is then expected to get the final results proportional to the integration over the dark matter momentum $\int_{\vec{k}}|\phi_{\vec{k}}|^2$ instead of $|\phi_0|^2$ obtained in the monochromatic approximation.

\section{$\nu \to f$ transition}

Let us first consider the scalar dark matter coupling to a neutrino and a (Dirac or Majorana) singlet fermion as a mediator which is heavier than the neutrino ($m_f > m_\nu$).
\begin{equation} \label{phinuf}
{\cal L}'= g \hat\phi \overline{ f_R} \nu_L + g^* \hat\phi^\dagger \overline{\nu_L} f_R .
\end{equation}
The   $\nu \to f $ transition amplitude in the background of $\phi_c$ is given by
\begin{equation} \label{Anuf}
{\cal A}_{\nu \to f} =
\langle \phi_c; f, \vec{p}_f | e^{-i H_0 t_2} U(t_2,t_1) e^{iH_0 t_1} | \nu, \vec{p}_\nu; \phi_c \rangle
\end{equation}
where $H_0$ is the Hamiltonian for the free fields and $U(t_2,t_1)=T e^{i\int^{t_2}_{t_1} dt\int d^3x {\cal L}'(x)}$ is the unitary time evolution operator in the interaction picture.  
Let us note that a state in the interaction  picture is represented by 
 $\Phi_I(t,\vec{x})= e^{iH_0 t} \Phi(0,\vec{x}) e^{-iH_0 t}$ \cite{peskin}, and thus
the transition amplitude (\ref{Anuf}) is written with respect to given initial and final states at a fixed time ($t=0$).  
The free Hamiltonial evolution factor $e^{i H_0 t}$ is irrelevant for the current discussion, but is responsible for the description of 
the standard neutrino oscillation process when $U=I$ as will be discussed in comparison with the new transition process 
in the next section.

 Up to the first order in perturbation, we have $U \sim  I + i \int\! d^4 x {\cal L}'(x)$. The zeroth order contribution with $U=I$  vanishes in this case, and the first order term in ${\cal L}'$ gives non-vanishing contribution to the amplitude ${\cal A}_{\nu \to f}$.
In this process, the momentum conservation  is enforced after the space integration $\int d^3 x \, e^{i(\vec{p}_f-\vec{p}_\nu \mp \vec{k}_0)}$ . Thus, the $\nu \to f$ transition probability obtained after the final state momentum integration is the sum of two absolute-squares: $P_{\nu\to f}=\int\!{ d^3\vec{p}_f \over (2\pi)^3 2E_f} [|{\cal A}^+_{\nu\to f}|^2+|{\cal A}^-_{\nu \to f}|^2]$ where
\begin{equation} \label{Apm}
\begin{split}
{\cal A}^{+}_{\nu\to f} &\propto  \int^{t_2}_{t_1}dt' e^{i(E_f-E_\nu - E_{\vec{k}_0})t'}
 (2\pi)^3 \delta^{(3)}(\vec{p}_f - \vec{p}_\nu - \vec{k}_0)   \\
&\left[
g \phi_0 \bar u_f(\vec{p}_{f},s)P_L u_\nu(\vec{p}_\nu, s) -
g^*{\phi}_0 \bar v_\nu(\vec{p}_\nu,s)P_R v_f(\vec{p}_f , s) \right]
\end{split}
\end{equation}
where the amplitude ${\cal A}^{+}$ involves $\phi_0 e^{-ik_0\cdot x}$ and the neutrino operator part $\nu_L(x)\sim u_L a_{\vec{p}_\nu} e^{-ip_\nu\cdot x}$ and  $\bar{\nu}_L(x) \sim \bar{v}_L a_{\vec{p}_\nu} e^{-i p_\nu \cdot x}$ for the first and second contribution, respectively. Then ${\cal A}^{-}_{\nu\to f}$ is obtained by the replacement: $\phi_0 \to \phi_0^*$ and $( E_{\vec{k}_0},  \vec{k}_0) \to -( E_{\vec{k}_0},  \vec{k}_0)$. Although trivial, let us also note that there is no interference between ${\cal A}^+$ and ${\cal A}^-$ as they correspond to transition amplitudes to different final states with $p_\nu \pm k_0$. 

 It is crucial to realize that a  non-trivial contribution, proportional to $\phi_0\neq 0$, arises as a medium effect in a coherent state which differs from an incoherent $N$-particle state $|N\rangle \propto (a^\dagger_{\vec{k}_0})^N|0\rangle$ leading to $\langle N| \hat \phi(x)|N\rangle =0$. 
Note that this process differs from the coherent forward scattering that involves no momentum transfer and thus modifies the fermion propagator in the background as depicted in Fig.~1. 
For these phenomena to happen,  incoherent scatterings need to be safely ignored so that the particles pass ``freely'' through a medium as in the case of solar neutrinos where the Wolfenstein effect is realized.
Indeed, one can find that the mean free path of neutrinos undergoing the interaction (2.1) with a ultra-light scalar medium  is extremely large in the most parameter space. This is nothing but the limits discussed in \cite{Choi:2019ixb} and will be shown explicitly in Figs.~2 and 3.

After taking  into account all the factors and normalization properly,  one obtains the transition probability which is the sum of four separate contributions for a Dirac fermion $f$:
\begin{equation}
\begin{split}
P_{\nu\to f} =& \left\{  \left| g \phi_0\int^{t_2}_{t_1}dt' e^{i(E_f-E_\nu -E_{\vec{k}_0} )t'} \right|^2
 {{1\over2} \sum_s |\bar u_f(\vec{p}_\nu+\vec{k}_0,s)P_L u_\nu(\vec{p}_\nu, s)|^2 \over 2 E_f 2 E_\nu}\right.
 \\
& +  \left.  \left| g^* \phi_0 \int^{t_2}_{t_1}dt' e^{i(E_f-E_\nu -E_{\vec{k}_0})t'} \right|^2
{{1\over2} \sum_s |\bar v_\nu(\vec{p}_\nu,s)P_R v_f(\vec{p}_\nu+\vec{k}_0, s)|^2 \over 2 E_f 2 E_\nu} \right\} \\
+& \left\{ \phi_0 \to \phi_0^*, ~(E_{\vec{k}_0},  \vec{k}_0) \to -( E_{\vec{k}_0},  \vec{k}_0) \right\}
\end{split}
\end{equation}
where the factor $2E_f 2E_\nu$ in the denominator appears due to the final state momentum integration and the state normalization $
\langle \vec{p}' | \vec{p}\rangle = (2\pi)^3 2E_{\vec{p}}\, \delta^3(\vec{p}'-\vec{p})$. 
In the practical situation of ultra-relativistic neutrinos propagating in the dark matter halo, we  have $E_f \approx E_{\nu} \approx |\vec{p}_{\nu}|$ and $(E_{\vec{k}_0},  \vec{k}_0) \approx (m_\phi,0)$ and thus 
\begin{equation}
P_{\nu\to f} =  \delta m^2  \left[{m_f^2 +m_\nu^2 \over 2E_\nu^2}\right]
\left[ {\sin^2\!\left(\!{\Delta_+ L\over2}\!\right) \over \Delta_+^2} + {\sin^2\!\left(\!{\Delta_- L\over2}\!\right) \over \Delta_-^2} \right]
\end{equation}
where   $\Delta_\pm \equiv {m_f^2 -m_\nu^2 \over 2 E_\nu} \pm m_\phi$ and $L=t_2-t_1$.
Here we define the medium-induced mass-squared $\delta m^2 \equiv 2|g|^2 |\phi_0|^2$.

In the limiting cases of  $m_\phi\ll m_f^2/2E_\nu = 5 \times 10^{-7} \mbox{eV} (m_f/\mbox{eV})^2/(E_\nu/\mbox{MeV})$, and  $m_\phi\gg m_f^2/2E_\nu$, one finds the simple forms of the transition probability:
\begin{equation} \label{Pnuf}
P_{\nu\to f} \approx \left\{
\begin{split}
& 4{ \delta m^2 \over m_f^2}
\sin^2\!\left(\!{m_f^2 L\over 4 E_\nu }\!\right) ~\mbox{for}~ m_\phi\ll m_f^2/2E_\nu , \\
&  {\delta m^2 m_f^2 \over  m_\phi^2 E_\nu^2}\sin^2\!\left(\!{m_\phi L\over2}\!\right) 
                  ~\mbox{for}~ m_\phi\gg m_f^2/2E_\nu .
\end{split}
\right.
\end{equation}
This transition has to be suppressed not to  reduce neutrino fluxes observed in various neutrino experiments. 
Considering solar neutrinos with the average traveling distance $L \approx 1/(1.3 \times 10^{-18} \mbox{eV})$ and typical 10\% uncertainties in the measurements, we obtain the upper bound  as follows:
\begin{equation}
|g| \lesssim \left\{ 
\begin{split}
&1.4\times 10^{-4} \left( {m_\phi \over 10^{-6}\mbox{eV}}\right) \left( {m_f \over \mbox{eV}} \right)
  ~\mbox{for}~ m_\phi\ll m_f^2/2E_\nu , \\
& 2.9\times10^{-4}  \left( {m_{\phi} \over 10^{-6} \mbox{eV}}\right)^2 
\left({E_\nu \over \mbox{MeV}}\right) \left( { \mbox{eV} \over m_f} \right) 
 ~\mbox{for}~ m_\phi\gg m_f^2/2E_\nu .
 \end{split} \right.
\end{equation}
%
Fig.~2 shows  the  bounds (black solid lines) from solar neutrinos with two different values of the mediator fermion mass $m_f=1$ eV and 100 eV.
Also shown are the limits from  the neutrino-dark matter scattering  assuming a real scalar dark matter \cite{Olivares-DelCampo:2017feq} and considering the SN1987A and IceCube-170922A observations \cite{Choi:2019ixb,Choi:2020ydp}. 
\begin{figure}
\centering
\includegraphics[width=0.46\textwidth]{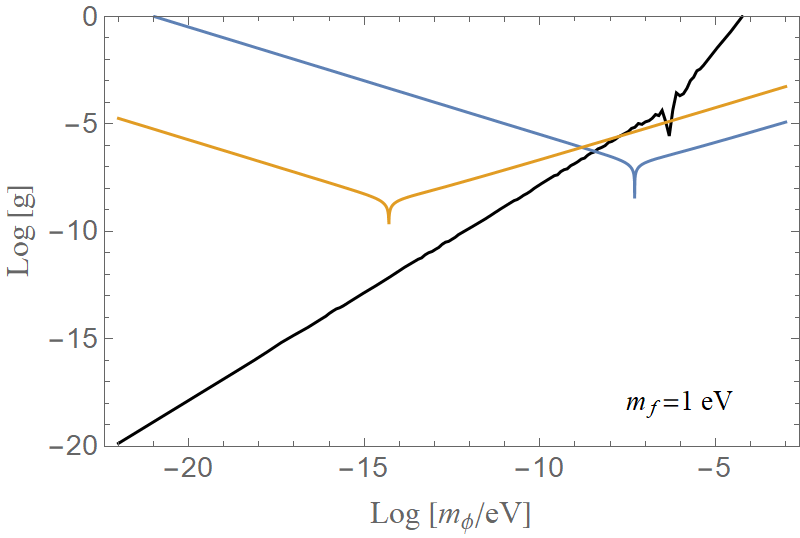}
\includegraphics[width=0.46\textwidth]{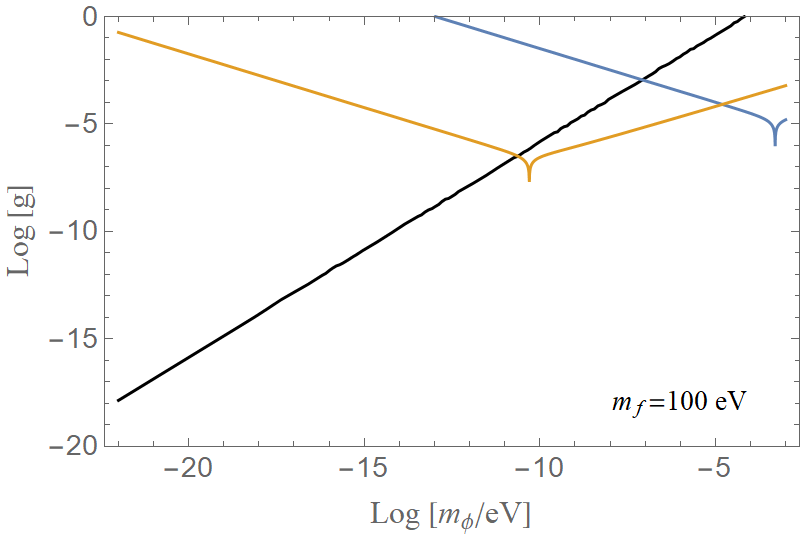}
\caption{The black solid lines show constraints on the coupling $g$ in terms of the dark matter mass $m_\phi$ considering  the solar neutrino transition. The orange and blue solid lines are from IceCube and SN1987A neutrino observations, respectively.}
\label{fig:mf1}
\end{figure}

\section{$\nu_\alpha \to \nu_\beta$ transition}

Let us now consider the ultra-light  scalar dark matter coupling to (Majorana) neutrinos:
\begin{equation}
{\cal L}'= {1\over 2}g_{\alpha\beta} \, \hat\phi \overline{\nu^c_{\beta R}} \nu_{\alpha L} 
+ {1\over 2} g_{\alpha\beta}^* \, \hat\phi^\dagger\, \overline{\nu_{\alpha L}}  \nu^c_{\beta R}
\end{equation}
written in the flavor basis with $\alpha, \beta=e, \mu$ and $\tau$.
For simplicity, we consider the two-flavor case and discuss the basic features of the 
neutrino transition in medium. Assuming that the medium-induced masses are much smaller than the bare mass terms, it is convenient to go to the  basis of  bare mass eigenstates.
The flavor eigenstates $\nu_{e,\mu}$ are expressed in terms of the mass eigenstates $\nu_{1,2}$ with the rotation angle $\theta$:
\begin{equation}
\begin{split}
\nu_e = & c_\theta \nu_1 - s_\theta \nu_2 \\
\nu_\mu = &  s_\theta \nu_1 +c_\theta \nu_2 .
\end{split}
\end{equation}
Then, the $\nu_e \to \nu_\mu$ transition amplitude is described by
\begin{equation} \label{Amue}
{\cal A}_{\mu e} = c_\theta s_\theta ({\cal A}_{11} - {\cal A}_{22}) 
   +c_\theta^2 {\cal A}_{21}-s_\theta^2 {\cal A}_{12} ,
\end{equation}
where the $\nu_i \to \nu_j$ transition amplitude in the mass basis ($i,j=1,2$) is calculated from 
\begin{equation}
{\cal A}_{ji} =
\langle \phi_c; \nu_j , \vec{p}_2 | e^{-i H_0 t_2} U(t_2,t_1) e^{iH_0 t_1} | \nu_i , \vec{p}_1; \phi_c \rangle 
\end{equation}
as described in the previous section.
One can notice that  the zero-th order contribution with $U=I$ describes the standard vacuum oscillation. 
Following the calculation in the previous section with the appropriate state normalization and 
the final state momentum integration, one obtains the oscillation probability which is expressed usually in terms of the ``effective'' amplitude ${\cal \tilde A}^0_{\mu e}$ defined by 
\begin{equation}
\begin{split}
P^0_{\mu e} &\equiv |{\cal \tilde A}^0_{\mu e}|^2=c^2_\theta s^2_\theta |{\cal \tilde A}^0_{11}- {\cal \tilde A}^0_{22}|^2 \\
\mbox{where} &~ {\cal \tilde A}^0_{ji} \equiv e^{-i E_i L} \delta_{ji}~~\mbox{with}~~ L=t_2- t_1,
\end{split}
\end{equation}
and thus the standard oscillation probability $P^0_{\mu e}=\sin^2{2\theta} \sin^2(\Delta m^2_{21} L/4E_\nu)$ is recovered.
It is useful to remind that this oscillation process can be affected by the medium effect modifying the dispersion relation of neutrinos [16,17]
which is assumed to be negligible in this calculation.

We are now ready to analyze the medium effect in the transition amplitude ${\cal A}_{ji}$. 
Repeating the previous calculation processes, one can obtain the effective transition amplitude ${\cal \tilde A}^\pm_{ji}$ given by
\begin{equation} \label{Aji}
{\cal \tilde A}^\pm_{ji}=e^{\pm i m_\phi (t_1 + {L\over2})} e^{-i {(E_j+E_i)L\over2} } \,
\sin{\!\left( \Delta^\pm_{ji} L \over2 \right)} 
{ \left[ g_{ji} \phi^{(*)}_0 \bar u_R(p_2) u_L(p_1) + g^*_{ji} \phi^{(*)}_0 \bar u_L(p_2) u_R (p_1) \right]
\over \Delta^\pm_{ji} E_\nu}
\end{equation}
where $\Delta^\pm_{ji} \equiv { \Delta m^2_{ji}\over 2E_\nu} \pm m_\phi$, $\vec{p}_2 = \vec{p}_1 \pm \vec{k}_0$, and $E_\nu \approx |\vec{p}_{1,2}|$ applicable to the usual neutrino oscillation phenomena.  Here  $\Delta m^2_{ji}$ is the mass-squared difference between two mass eigenstates $\nu_j$ and $\nu_i$.  
As discussed in the previous section, ${\cal \tilde A}^\pm_{ji}$ correspond  to the amplitudes of two different processes $\nu_i(\vec{p}_1) \to \nu_j(\vec{p}_1\pm \vec{k}_0)$.  Now the amplitude ${\cal A}_{\mu e}$ in (\ref{Amue})  is seperated by three independent  contributions: ${\cal A}_{\mu e}^0$ describing the standard oscillattion $\nu_e(\vec{p}) \to \nu_\mu(\vec{p})$, 
and ${\cal A}_{\mu e}^\pm $ describing the medium-induced up and down transition  $\nu_e(\vec{p}) \to \nu_\mu(\vec{p}\pm \vec{k}_0)$.
Therefore, the total transition probability receives three independent contributions: 
\begin{equation}
P_{\mu e}= |{\cal \tilde A}^0_{\mu e}|^2 +  \overline{|{\cal \tilde A}^+_{\mu e}|^2} 
+  \overline{|{\cal \tilde A}^-_{\mu e}|^2}
\end{equation}
where the spin average for $|{\cal A}^\pm_{\mu e}|^2$ is performed.  
Notice that the time-dependent factor in (\ref{Aji}) is an overall phase, and thus disappears in the transition probability.

In the bare mass basis, there can appear generic couplings of all components $g_{ij}$ which makes highly nontrivial the fully consistent treatment.
In this analysis, we assume the presence or dominance of only one coupling, $g_{22}$ or $g_{12}$, to see how the neutrino transition behaves
in a medium.  It is then found that the new contributions to the transition probability in each case are given by 
\begin{eqnarray}
 \overline{|{\cal \tilde A}^\pm_{\mu e}|^2} & =& \sin^22\theta\,
  \frac{\delta m^2  m^2_{\nu_2}}{2 m_\phi^2 E_\nu^2}\, \sin^2 { m_\phi L\over2}
~~~~~\mbox{for}~ \delta m^2= 2 |g_{22}|^2 |\phi_0|^2 ,  \label{eq:P22}\\ 
 \overline{|{\cal \tilde A}^\pm_{\mu e}|^2}  & = & 2 \cos^22\theta\,
  {\delta m^2  m_{\nu_2}^2 \over (\Delta^\pm_{21} E_\nu)^2 }\, \sin^2 { \Delta_{21}^\pm L\over2}
~~\mbox{for}~ \delta m^2= 2 |g_{12}|^2 |\phi_0|^2 , \label{eq:P12}
\end{eqnarray}
where  $m_{\nu_2} \gg m_{\nu_1}$ is assumed for the second case.
 
 \begin{figure}
\centering
\includegraphics[width=0.6\textwidth]{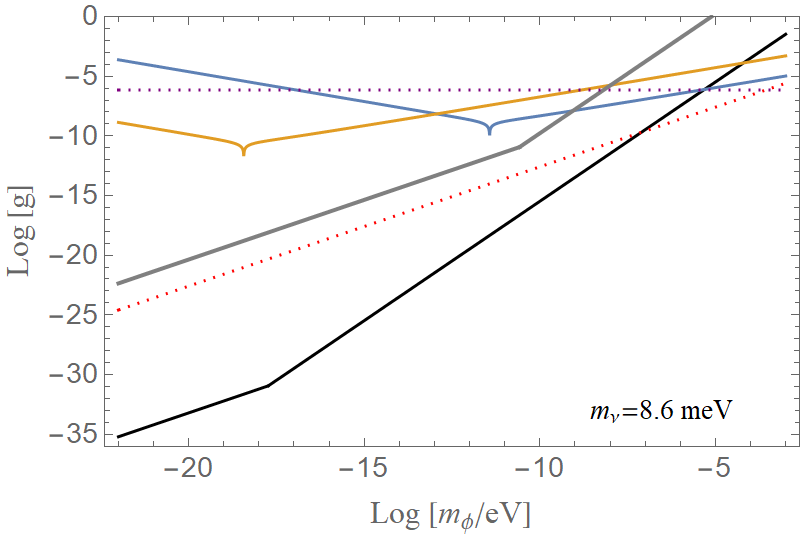}
\caption{The constraints on the coupling $g_{22}$ (lower black solid line) or $g_{12}$ (upper gray solid line) from  the solar neutrino transition. The orange and blue solid lines are from IceCube and SN1987A neutrino observations, respectively. The upper purple dotted and lower red dotted lines show the CMB and BBN constraints, respectively.  }
\label{fig:nusol}
\end{figure}
 
Considering the solar neutrino oscillations, we take $s_\theta^2=0.31$, $\Delta m^2_{21} = 7.4 \times 10^{-5} \mbox{eV}^2$, $E_\nu = 1$ MeV, 
and $L=1/(1.3 \times 10^{-18} \mbox{eV})$ \cite{pdg} to get
a very stringent upper limit  for the coupling $g_{22}$ with two distinctive $m_\phi$ dependencies:
 \begin{equation} \label{eq:g22}
| g_{22}| \lesssim \mbox{Max}\left[ 6.7\times 10^{-32} \left( m_\phi \over 10^{-18} \mbox{eV}\right), \, 
 3.4\times 10^{-32} \left( m_\phi \over 10^{-18} \mbox{eV}\right)^2  \right].
 \end{equation}
 On the other hand, a much weaker bound is obtained for $g_{12}$:
  \begin{equation} \label{eq:g12}
| g_{12}| \lesssim \mbox{Max}\left[ 7.2\times 10^{-12} \left( m_\phi \over 10^{-11} \mbox{eV}\right), 
 2.7\times 10^{-12} \left( m_\phi \over 10^{-11} \mbox{eV}\right)^2  \right]
 \end{equation}
applicable to two regions of $m_\phi$ separated by  $ \Delta m^2_{21}/2E_\nu \approx 3.7\times 10^{-11}$ eV with $E_\nu=1$ MeV.
 
 In Fig.~3, the lower black and upper gray solid lines show the bounds  from (\ref{eq:g22}) and  (\ref{eq:g12}), respectively. 
Also shown are the upper limits on the neutrino-dark matter  interactions obtained from SN1987A and IceCube-170922A observations.
 Whether or not the light scalar composes dark matter, one can draw constraints on its coupling to neutrinos through the processes mediated by the scalar field. The horizontal purple-dotted line is from the neutrino-neutrino scattering effect in the CMB measurement \cite{Forastieri:2019cuf}, and the diagonal red-dotted line is from its impact on the effective number of neutrinos during BBN \cite{Venzor:2020ova} applied to $g_{22}$ with $m_{\nu_2}=8.6$ meV. 
In a generic situation, we expect that the stronger bound on $g_{22}$ is applicable to the Yukawa couplings $g_{\alpha\beta}$ up to some mixing angle dependencies.

\section{Medium-induced neutrino oscillations}

Let us finally comment on the cases where the neutrino oscillations are solely due to the medium effect [16,17]. 
In this case, the (approximate) mass basis is the diagonal basis of the couplings $g$. Then, the neutrino transition is described by the second equation of (\ref{Pnuf}), 
or the equation for the diagonal component (\ref{eq:P22}).  Considering the solar neutrino oscillation again, one obtains 
the bounds on the bare masses $m_f$ and $m_\nu$ as follows:
\begin{eqnarray}
{m_f \over \mbox{eV}}  &<&  \mbox{Max}\left[0.96\times 10^{-10},\,  5.2\times 10^{-3} {m_\phi \over 10^{-10} \mbox{eV}}\right] , \\
{m_\nu \over \mbox{eV} } &<&  \mbox{Max}\left[1.4\times 10^{-10},\,  7.3\times 10^{-3} {m_\phi \over 10^{-10} \mbox{eV}}\right] ,
\end{eqnarray}
for $E_\nu=1$ MeV.
Here the upper limit of $m_\phi \lesssim 10^{-10}$ eV is imposed to satisfy the astrophysical bound from the SN1987A observation [17].

\section{Conclusion}

We have analyzed a new phenomenon of the neutrino transition in a medium of coherently oscillating scalar dark matter.
This occurs at the first order in perturbation of the neutrino-scalar interaction which describes the propagation of neutrinos exchanging tiny momentum with the medium. It is distinguished from the usual neutrino oscillations in a medium caused by the phase difference (or mass-squared difference) between two mass eigenstates.  Considering its impact on solar neutrinos,  stringent constraints on the model parameters are obtained and compared with the existing limits in a wide range of the dark matter mass $(10^{-22}\sim 10^{-3})$ eV.


\end{document}